\documentclass[]{spie}  

\usepackage{amsmath,amsfonts,amssymb}
\usepackage{graphicx}
\usepackage{subfig}
\usepackage[colorlinks=true, allcolors=blue]{hyperref}
\usepackage{siunitx}
\usepackage{float}
\DeclareSIUnit\angstrom{\text{Å}}
\DeclareSIUnit\photon{\text{ph}}
\DeclareSIUnit\arcsec{\text{arcsec}}
\usepackage{amssymb}
\usepackage{xcolor}

\title{Dynamic scheduling for SOXS instrument: environment,  algorithms and development}

\author[a,b]{Laura Asquini}
\author[a,c]{Marco Landoni}
\author[d]{Dave Young}
\author[e]{Laurent Marty}

\author[d]{Stephen J. Smartt}

\author[a]{Sergio~Campana}
\author[f]{Riccardo~Claudi}
\author[e]{Pietro~Schipani}
\author[a]{Matteo~Aliverti}
\author[f]{Federico Battaini}
\author[f]{Andrea~Baruffolo}
\author[g]{Sagi~Ben-Ami}
\author[a]{Andrea Bianco}
\author[h]{Federico~Biondi}
\author[e]{Giulio~Capasso}
\author[k]{Rosario~Cosentino}
\author[i]{Francesco~D'Alessio}
\author[a]{Paolo~D'Avanzo}
\author[g]{Ofir	Hershko}
\author[m]{Hanindyo~Kuncarayakti}
\author[k]{Matteo~Munari}
\author[n]{Giuliano~Pignata}
\author[g]{Adam~Rubin}
\author[k]{Salvatore~Scuderi}
\author[i]{Fabrizio~Vitali}
\author[l]{Jani~Achrén}
\author[n]{José~Antonio~Araiza-Duran}
\author[o]{Iair~Arcavi}
\author[p]{Anna~Brucalassi}
\author[g]{Rachel~Bruch}
\author[f]{Enrico~Cappellaro}
\author[e]{Mirko~Colapietro}
\author[e]{Massimo~Della~Valle}
\author[e]{Marco~De~Pascale}
\author[k]{Rosario~Di~Benedetto}
\author[e]{Sergio~D'Orsi}
\author[g]{Avishay~Gal-Yam}
\author[a]{Matteo~Genoni}
\author[q]{Marcos~Hernandez}
\author[m]{Jari~Kotilainen}
\author[i]{Gianluca~Li~Causi}
\author[m]{Seppo~Mattila}
\author[a]{Giorgio Pariani}
\author[g]{Michael~Rappaport}
\author[f]{Kalyan~Radhakrishnan}
\author[f]{Davide~Ricci}
\author[a]{Marco~Riva}
\author[f]{Bernardo~Salasnich}
\author[f]{Ricardo~Zanmar~Sanchez}
\author[r]{Maximilian~Stritzinger}
\author[q]{Hector~Ventura}

\affil[a]{INAF – Osservatorio Astronomico di Brera-Merate, via E. Bianchi 46, I-23807 Merate (LC), Italy;}
\affil[b]{Dipartimento di Scienza e Alta Tecnologia, Università dell’Insubria, via Valleggio 11, I-22100 Como, Italy}
\affil[c]{INAF - Osservatorio Astronomico di Cagliari. Via della Scienza 5, Selargius (CA) - Italy}
\affil[d]{Astrophysics Research Centre, School of Mathematics and Physics, Queen's University Belfast, Belfast BT7 1NN, UK}
\affil[e]{INAF - Osservatorio Astronomico di Capodimonte, Salita Moiariello 16, Naples- Italy }
\affil[f]{INAF -- Osservatorio Astronomico di Padova, Vicolo dell’Osservatorio 5, I-35122, Padua, Italy }

\affil[g]{Weizmann Institute of Science, Herzl St 234, Rehovot, 7610001, Israel }
\affil[h]{Max-Planck-Institut für Extraterrestrische Physik, Giessenbachstr. 1, D-85748 Garching, Germany }
\affil[k]{INAF -- Osservatorio Astrofisico di Catania, Via S. Sofia 78 30, I-95123 Catania, Italy }
\affil[i]{INAF -- Osservatorio Astronomico di Roma, Via Frascati 33, I-00078 M. Porzio Catone, Italy }
\affil[l]{Incident Angle Oy, Capsiankatu 4 A 29, FI-20320 Turku, Finland }
\affil[m]{Finnish Centre for Astronomy with ESO (FINCA), FI-20014 University of Turku, Finland}
\affil[n]{Universidad Andres Bello, Avda. Republica 252, Santiago, Chile }
\affil[o]{Tel Aviv University, Department of Astrophysics, 69978 Tel Aviv, Israel }
\affil[p]{INAF - Osservatorio Astrofisico di Arcetri.Largo Enrico Fermi 5, 50125 Florence - ITALY }

\affil[q]{FGG-INAF, TNG, Rambla J.A. Fernández Pérez 7, E-38712 Breña Baja (TF), Spain }
\affil[r]{Aarhus University, Ny Munkegade 120, D-8000 Aarhus, Denmark }
\authorinfo{laura.asquini@inaf.it}
\begin{document} 
\maketitle

\begin{abstract}
We present development progress of the scheduler for the Son Of X-Shooter (SOXS) instrument at the ESO-NTT 3.58 meter telescope. SOXS will be a single object spectroscopic facility, consisting of a two-arms high-efficiency spectrograph covering the spectral range 350-2000 nanometer with a mean resolving power R$\approx$4500. SOXS will be uniquely dedicated to the UV-visible and near infrared follow up of astrophysical transients, with a very wide pool of targets available from the streaming services of wide-field telescopes, current and future. This instrument will serve a variety of scientific scopes in the astrophysical community, with each scope eliciting its specific requirements for observation planning, that the observing scheduler has to meet. Due to directions from the European Southern Observatory (ESO), the instrument will be operated only by La Silla staff, with no astronomer present on the mountain. This implies a new challenge for the scheduling process, requiring a fully automated algorithm that should be able to present the operator not only with and ordered list of optimal targets, but also with optimal back-ups, should anything in the observing conditions change. This imposes a fast-response capability to the scheduler, without compromising the optimization process, that ensures good quality of the observations. In this paper we present the current state of the scheduler, that is now almost complete, and of its web interface. 
\end{abstract}

\newpage
\section{INTRODUCTION}
\label{sec:intro}  
The Son Of X-Shooter (SOXS \cite{schipani2016new, schipani2018soxs}) instrument is a single-object spectrograph with spectral resolution R$\approx$4,500 that will observe the southern sky from the New Technology Telescope (NTT) in La Silla, Chile. It will be operated in the ESO La Silla-Paranal Observatory operation environment, without an astronomer on the mountain. From this stems the need to design and develop a web-based software that can autonomously organize and manage the night, both in advance (for scientists to approve and scrutinize) and on-the-fly.  Both these processes must ensure the respect and optimization of observational constraints, as well as a periodic check on the current weather on the mountain. 
Further, the scheduler must combine targets from the SOXS Consortium and regular ESO proposals, respecting the 50/50 Guaranteed Time of Observation agreed upon. These two entities will feed the scheduler through different channels, namely the Marshall application (see x) for follow-up observations of Targets of Opportunity (ToO) for the Consortium and the Phase 2 (P2) web page for other ESO users. During and at the end of the night, the scheduler must be able to keep track of the operations of the instrument each night in terms of percentage of completed observations, hit/miss rate of transient events and general observing conditions for each target. The scheduler will communicate automatically with the official ESO Phase 2 tools, creating and verifying Observing Blocks (OBs) from the targets in the standard ESO way, sending them to fill the Visitor Execution Sequence (VES) according to the visibility of the target, their priority and current weather condition (accessed through the Astronomical Site Monitor - ASM - Weather Application Programming Interface - API). This paper is organized as follows:
In Section \ref{sec:sched} we will introduce an overall view of the scheduler and its workflow; in Section \ref{sec:marshall} we will briefly recall the functioning of the Marshall application, and describe its interaction with the SOXS local database; in Section \ref{sec:filter} we will recall the filtering strategies for the targets the scheduling algorithms, with particular emphasis on the new features available. Section \ref{sec:nightw} will explain the current architecture of the night management, and Section \ref{interface} will conclude with a description of the Web interface. 

\section{Marshall and database}
\label{sec:marshall}
The Marshall Web Application\cite{landoni2020soxs} for the SOXS instrument is an adaptation of the currently available Marshall developed for the ESO extended Public ESO Spectroscopic Survey of Transient Objects (ePESSTO\cite{smartt2015pessto}). This application is in charge of collecting all data coming from various survey streaming services and, for each target, aggregating them in a single, uniquely identified object. This is done by producing a “ticket” with a unique numeric ID that the scheduler can use as an identifier for that object. Further, the Marshall keeps track of the evolution of the target and keeps the scheduler local database updated on variations (magnitude, time stamps, classification, comments). Further, the Marshall will be in charge of creating and sending to the local database Follow-Up OBs. These targets will be completely customized in each of their aspects, from exposure to instrument mode. The local SQL database receives the targets and creates an OB that is uniquely identified by its ID, and contains all of the information needed for observation and constraint evaluation, such as right ascension, declination, magnitude, and so on. 
The data are then stored and processed by the filtering scripts.  

\section{The SOXS Scheduler}
\label{sec:sched}
 The scheduler is designed as a RESTful API using the Flask micro-framework that acts as mediator between the Marshall and the ESO P2 system. The OBs are created through the SOXS scheduler and are stored in a standard MySQL database. The main part of the code is developed so that any ground-based instrument could in principle use this scheduler, with only two additional files containing all of the specifics and default values of the instrument and the telescope.
\begin{figure} [h!]
    \centering
    \includegraphics[width=0.75\textwidth]{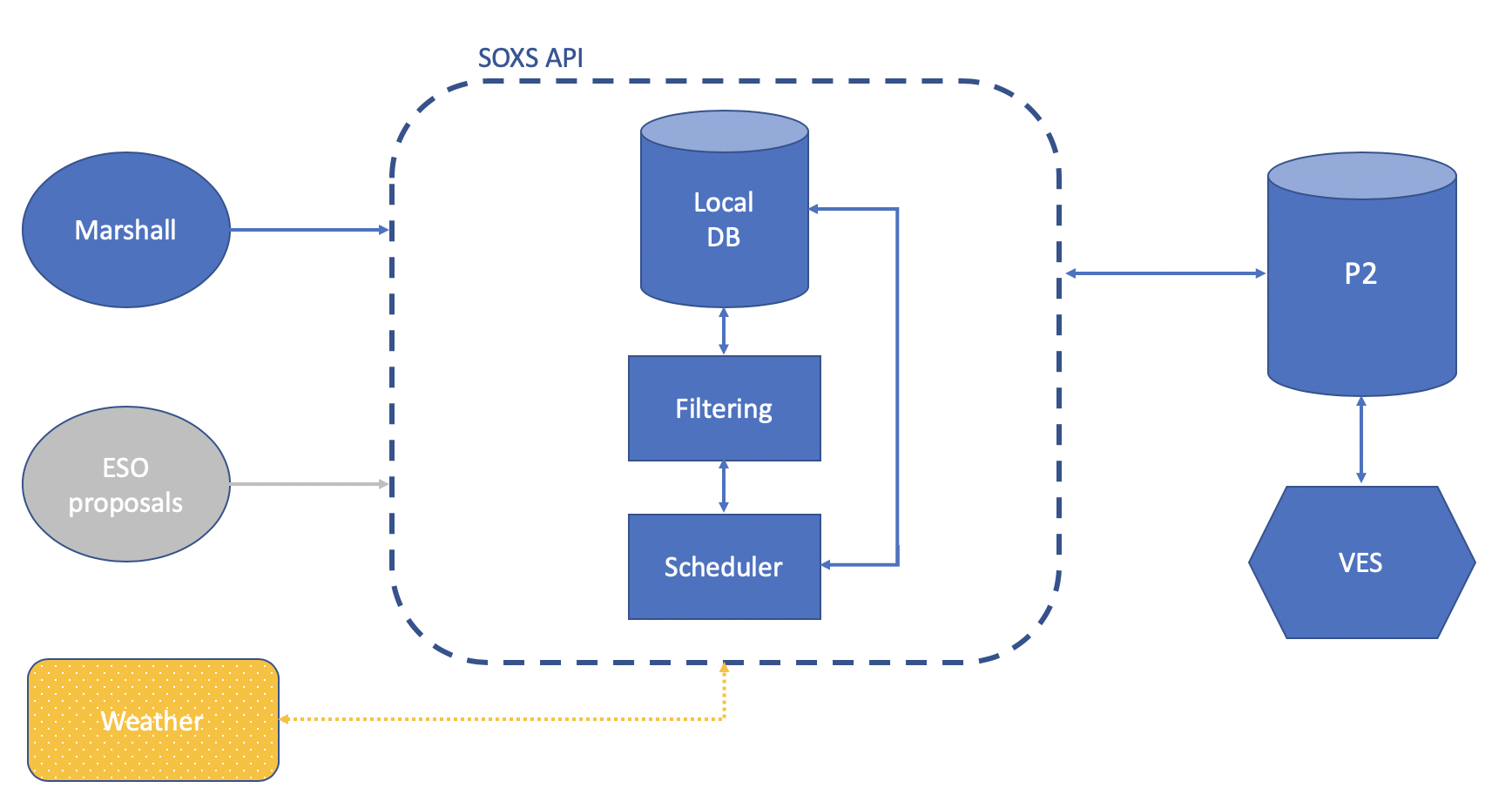}
    \caption{Overall workflow of the scheduler. Blue blocks mean that the relative part is working and running, while the yellow block is currently being worked-on. Gray blocks indicate that no testing will be done on that object for now.}
    \label{fig:overallflow}
\end{figure}
The scheduling algorithm is developed in Python 3 and uses tools from Astropy \cite{robitaille2013astropy, price2018astropy} libraries, in particular \cite{Astroplanmorris2018astroplan}. Astroplan is a tool designed for observation planning, and it provides with common observational quantities such as celestial rising and setting times at a given Earth location. For our purposes, this tool is used to compute observational constraints for the targets while also employing its built-in priority scheduler. In order to ensure fast response and updates, we divided the algorithm’s workflow into two steps. The first step handles most of the computational workload and selects targets that can be properly observed during a designated night. This is done by means of two scripts, one that processes newly discovered transients, and another one that updates the already registered ones. The second step is the actual scheduling process. 
As shown in fig. \ref{fig:overallflow}, the data stream begins with either the Marshall or ESO P2 feeding the SOXS API with targets. These targets are ingested by the database, and are processed daily by the filtering algorithms, that ensure their information is up-to-date and ready to be scheduled, if visible. The scheduler then creates a schedule (more on this in Section \ref{sec:filter}) optimizing both the observational constrains and the use of the night. Once the schedule is approved, the scheduler sends each OB one at the P2 API, validates it, and sends it to the VES. From now on, a constant feedback from P2 is requested, with the status of the observations. Meanwhile, a query is periodically sent to the ASM API in order to check for weather conditions. If all goes well, the status of the OB is updated both in the local database and on the P2, and the next OB is sent out. 
On the other hand, should anything go wrong, the automatic night manager will solve the situation either by selecting another OB, or by rearranging the left-over schedule, but more on that in Section \ref{sec:nightw}.

\section{Filtering and scheduling}
\label{sec:filter}
\subsection{Filtering}
The filtering of targets is mandatory in order to reduce the workload on the scheduler. We have developed two scripts, one for processing new OBs and one for updating the pre-existing ones. These scripts will work constantly during day and night, periodically searching the database for their targeted OBs. The algorithm, which we described in detail in a previous paper\cite{landoni2020soxs}, is based on a representation of the night with 1D arrays where each slot represents a five-minute time span. An equal array, the observability array, is associated to each target, and the result of the convolution of all the observing constraints is represented either by a zero or a one in each slot, with zero meaning the object is not observable, and one that is is. Further, at this step we associate a grade to each target, that represents the likelihood of a good observation during that night. These grades are an adaptation of the currently used system used by ESO to plan observations\cite{bierwirth2010new}.
The processing script evaluates each object for five consecutive nights, while the updating script discards the past night and computes the fifth from the current day. This is done so that there is greater availability of targets, both for the scheduler and for the scientific team, at all times. 
\begin{figure} [h!]
    \centering
    \includegraphics[width=0.75\textwidth]{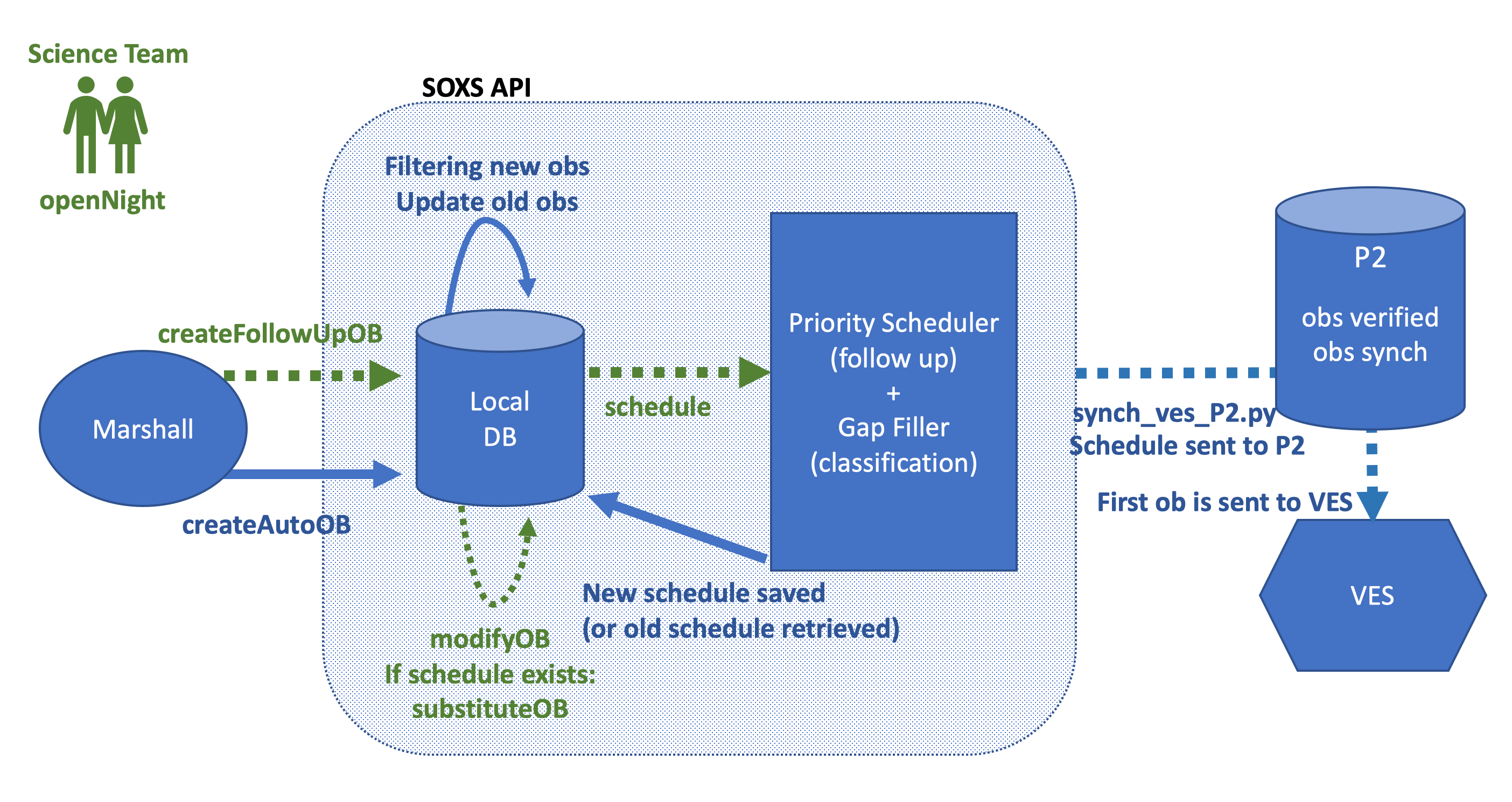}
    \caption{Simplified workflow of the scheduler, with action taken by the science team.}
    \label{fig:sciencet}
\end{figure}
\subsection{Scheduling}
\begin{figure} [h!]
    \centering
    \includegraphics[width=0.75\textwidth]{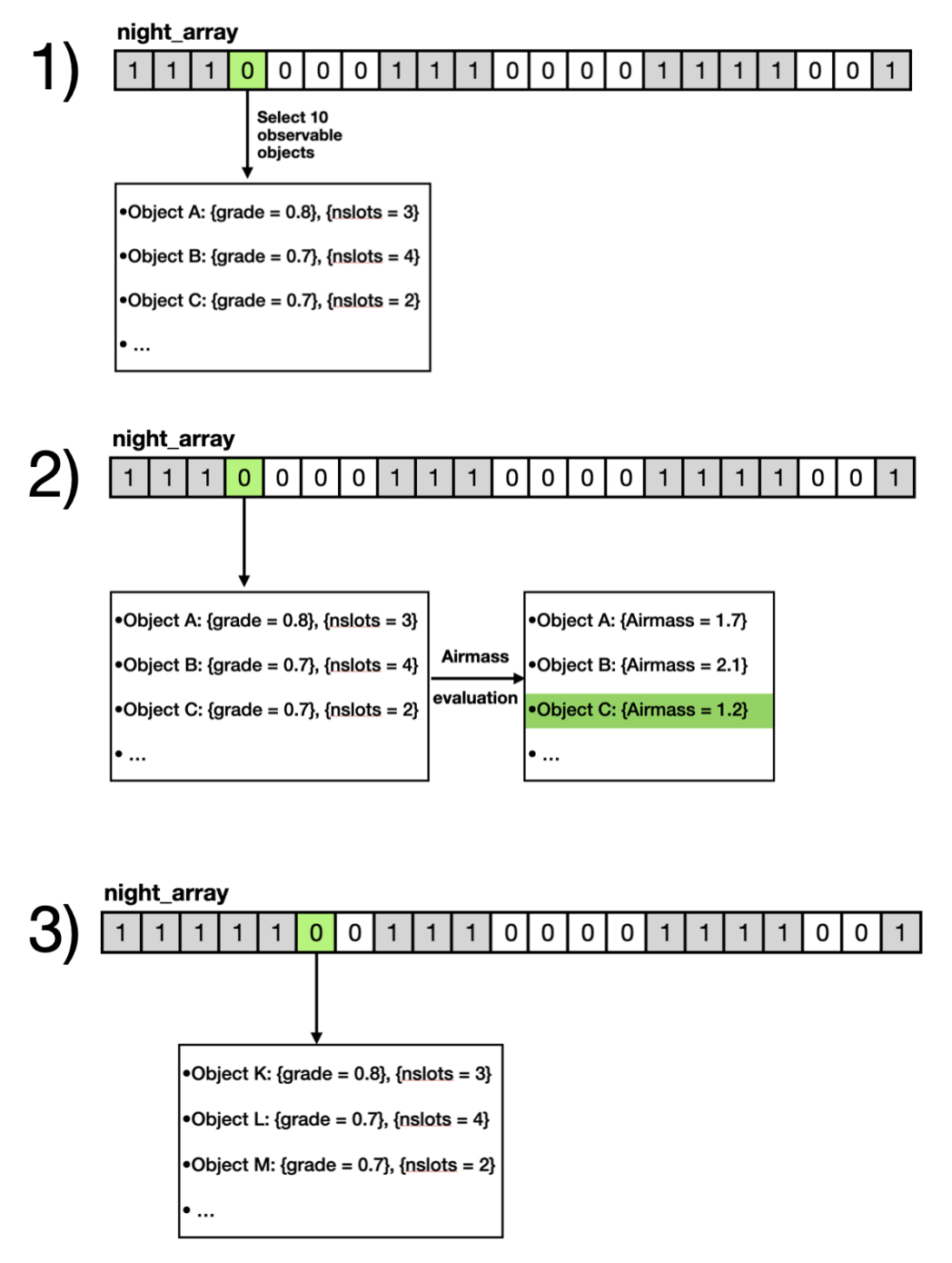}
    \caption{Visualization of the algorithm for optimized gap filling. Figure from  \cite{landoni2020soxs}.}
    \label{fig:gapfiller}
\end{figure}

In order to deal with the scientific prioritization of the targets, which is especially important for follow-up targets, and ESO proposals, the built-in Astroplan priority scheduler is implemented as a first step and deals with the top priority targets. This scheduler is designed to find the optimal observing time for each object, processing them based on their priority. The algorithm evaluates the constraints for each one with the very fine time resolution of one minute. Although this process is very efficient in optimizing the observing conditions, it is computationally quite demanding because of the time resolution at which the constraints are evaluated, hence the decision to employ this scheduler only for top priority objects. The task now remains to optimally make use of
the remaining time during the night. This is when our fixed parametrization of the night with 1D boolean arrays comes into play. Each target previously scheduled is assigned to the corresponding five-minute time span slot during the night. What the scheduler has now to deal with is a partially full night array, with many gaps to fill. In order to do so,
the algorithm looks for the first free slot in the night array and finds in the list of observable obs (sorted by their grades, with most likely successful objects on top) the first objects that fit the gap, by checking their observability arrays and slots required by the exposure time (we fix the number of fitting objects equal to 10). Having found such objects, the algorithm evaluates the constraints at the mean time of the candidate observation and chooses the best one. The winner of this competition is added to the schedule, and the night array is changed accordingly. This process is repeated until the night is full, or there are no objects in the list that can fill the gaps. A visualization of the workflow of this algorithm can be seen in Fig. \ref{fig:gapfiller}. At the end, the algorithm as produced an ordered list of objects to observe that night, each with their specified priority and with a specified observing time.

During an observing night, the scientific team will not have access to the algorithms nor they will need to. However, they will have the possibility to tweak and change the schedule to their liking, aided by the scheduler. At the beginning of the night, or whenever the scientific team convenes to approve a schedule, a new night is opened through the web interface. As long as the night is open, the scheduler and/or the scientific team can interact with the schedule. The scientific team then asks the scheduler to propose a schedule for the night, which the scheduler will produce with the optimized algorithm described in the previous paper\cite{landoni2020soxs}, respecting scientific priorities and observation-quality constraints. At this point, the scientific team can select an OB to substitute: the scheduler will provide alternatives, and make the swap upon selection. Not all obs can be swapped though, as for Follow-Up obs the possibility of selecting a specific time and date is provided, with a ``fixed" option, that will prevent any swapping or any scheduling outside the required time. Another way the scientific team can produce a schedule is by manually adding the obs. This is done by scanning the ``observable obs" page, and selecting the tagets sequentially. The selected OB will be checked for observability, and counter will provide the scientist with the information about the number of slots left in the night. Once the schedule is produced and possibly modified, the science team will approve it, thus synchronizing the contained targets with P2, where they are verified. The first OB of the schedule is then sent to populate the VES. An ideal workflow for the night is shown in Fig. \ref{fig:sciencet}

\section{Night Management}
\label{sec:nightw}
The managing of the night is the trickiest part to implement. This is not currently implemented, although the design part is already completed. The absence of an astronomer on the mountain implies that the scheduler must make fast and sensible decisions, in order not to waste any time at the telescope. There are many kinds of changes or events that can trigger the intervention of the scheduler, most importantly the arrival of a urgent ob, that is mandatory to observe and disrupts the current schedule. Besides, much more frequent events can occur, such as grave delays to observations (maybe due to instrument re-calibration or technical problems, which are not uncommon), change in wind direction and consequently of pointing restriction, changes in seeing conditions, cloud coverage or even in the target magnitude itself. This is why we chose to perform everyone of these checks at the end of the observation for every ob, calling this operation ``Fetch ob" (see Fig. \ref{fig:overallflow}. 
\begin{figure} [h!]
    \centering
    \includegraphics[width=0.75\textwidth]{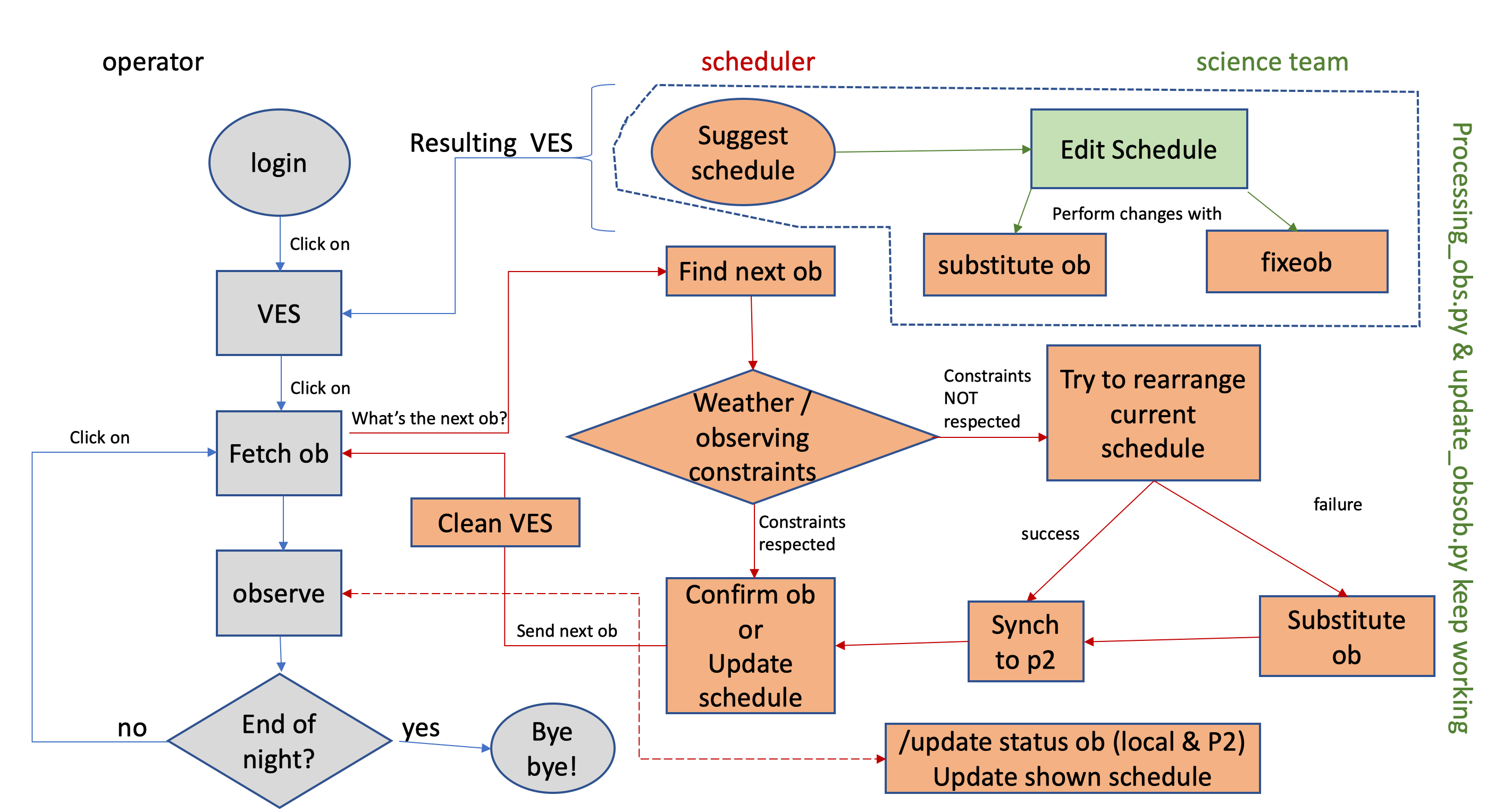}
    \caption{Workflow for the night of every actor involved. In gray, the actions performed by the operator on the mountain, in orange, those automatically performed by the scheduler, in green those involving the science team.}
    \label{fig:overallflow}
\end{figure}
The Fetch OB is called by the telescope operator every time an OB is about to finish, or at the beginning of the night. This will trigger the search for the next target in the schedule, checking on the OB status (immediately updated upon observation and upon its ending). The next OB will be then scrutinized: its magnitude will be checked for updates and a query to the Weather API will inform about the current weather conditions (clouds, wind and such) which will be compared to those required by the target. Further, a control on the current time and the supposed time of observation will be performed. If all goes well, the OB is cleared for observation and sent to the operator on the mountain. Otherwise, many things could in principle happen. Our line of action is to preserve as much of the schedule as possible, so that the scheduler will run the scheduling algorithms on the remaining time of the night, with the remaining obs. Given that these algorithms have particular attention to the scientific priority of targets, the most important ones should be saved and re-allocated at their optimum with the time available. If after this operation any gap remains, the scheduler will run once more with all of the obs in the database filling the empty time. Any changes are promptly synchronized and verified with P2, and updates on the OB status will be sent both to the local and to the P2 database. Previous discarded obs, however, will be checked for observability later in the night, also to stay as conservative as possible with the approved schedule. 

\section{Interface}
\label{interface}
\begin{figure}[h!]
    \centering
        \includegraphics[width=0.8\textwidth]{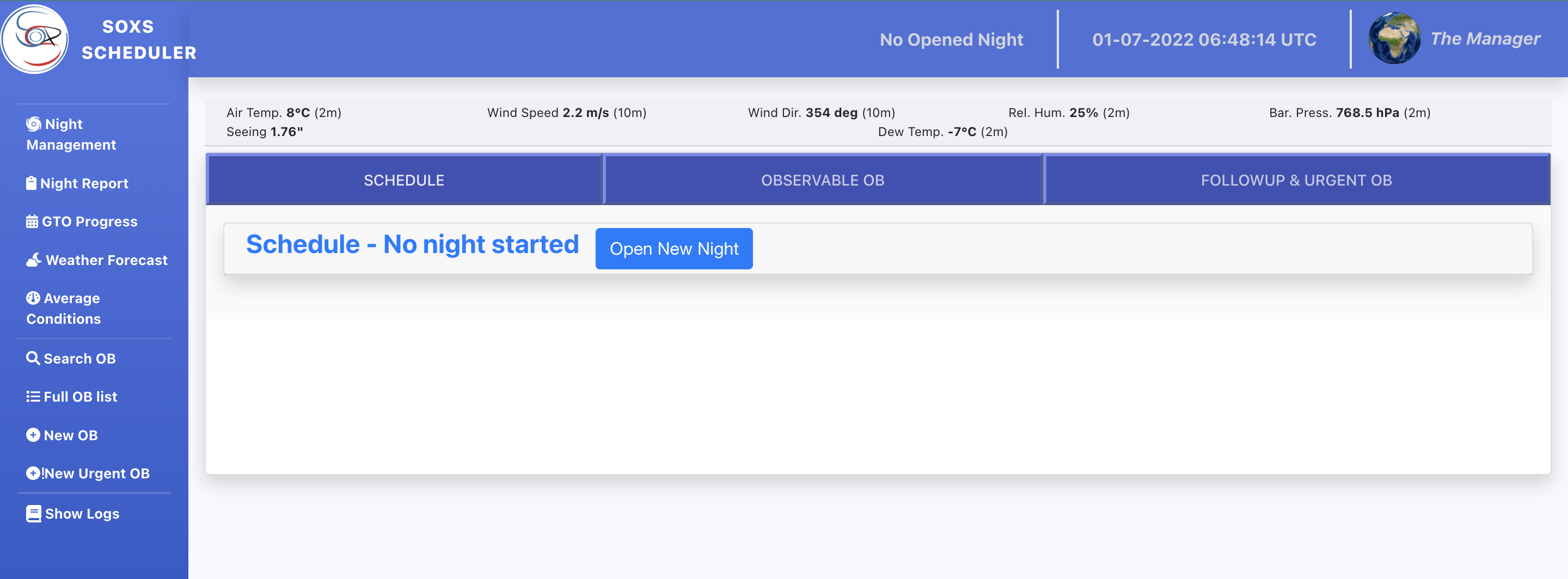}
        \includegraphics[width=0.8\textwidth]{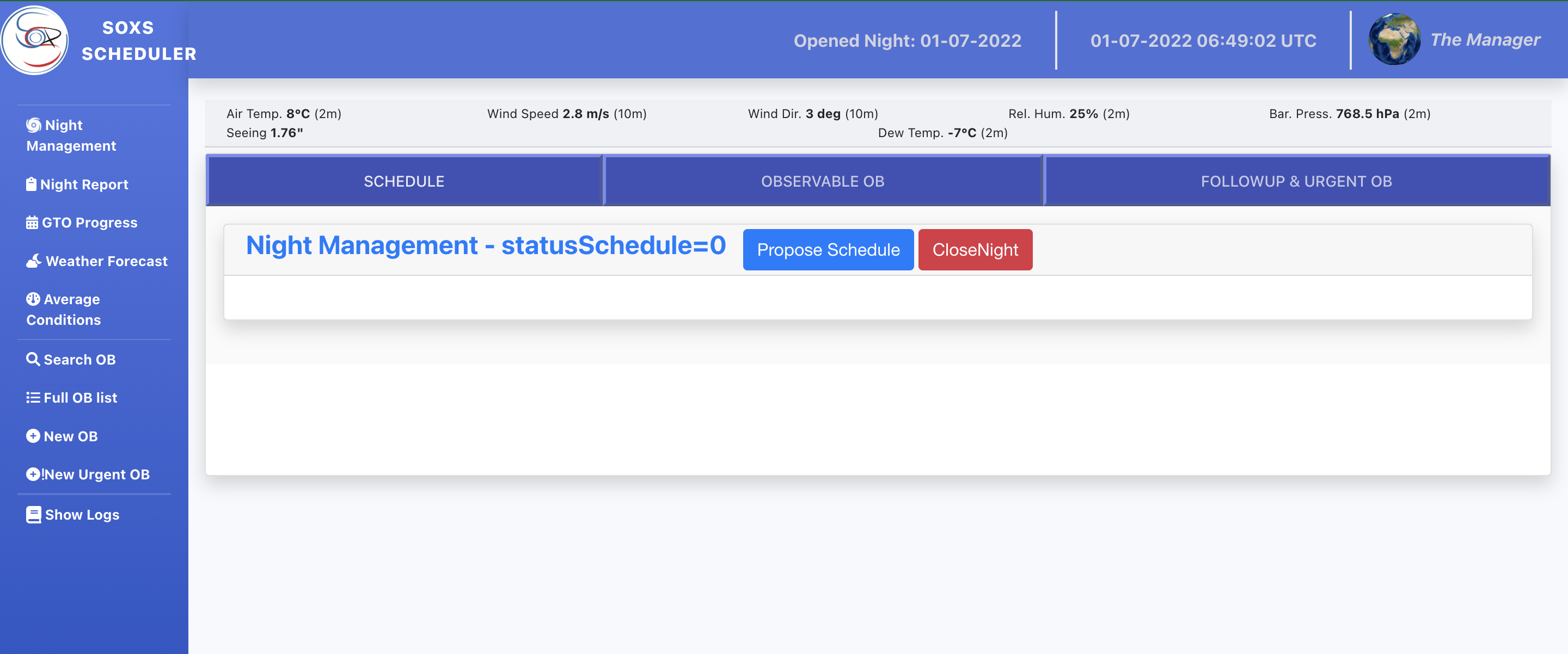}
    \caption{Top: home page of the Science Team interface upon login. Bottom: home page after opening a new night.}
    \label{fig:if01}
\end{figure}

\begin{figure}[h!]
    \centering
        \includegraphics[width=0.8\textwidth]{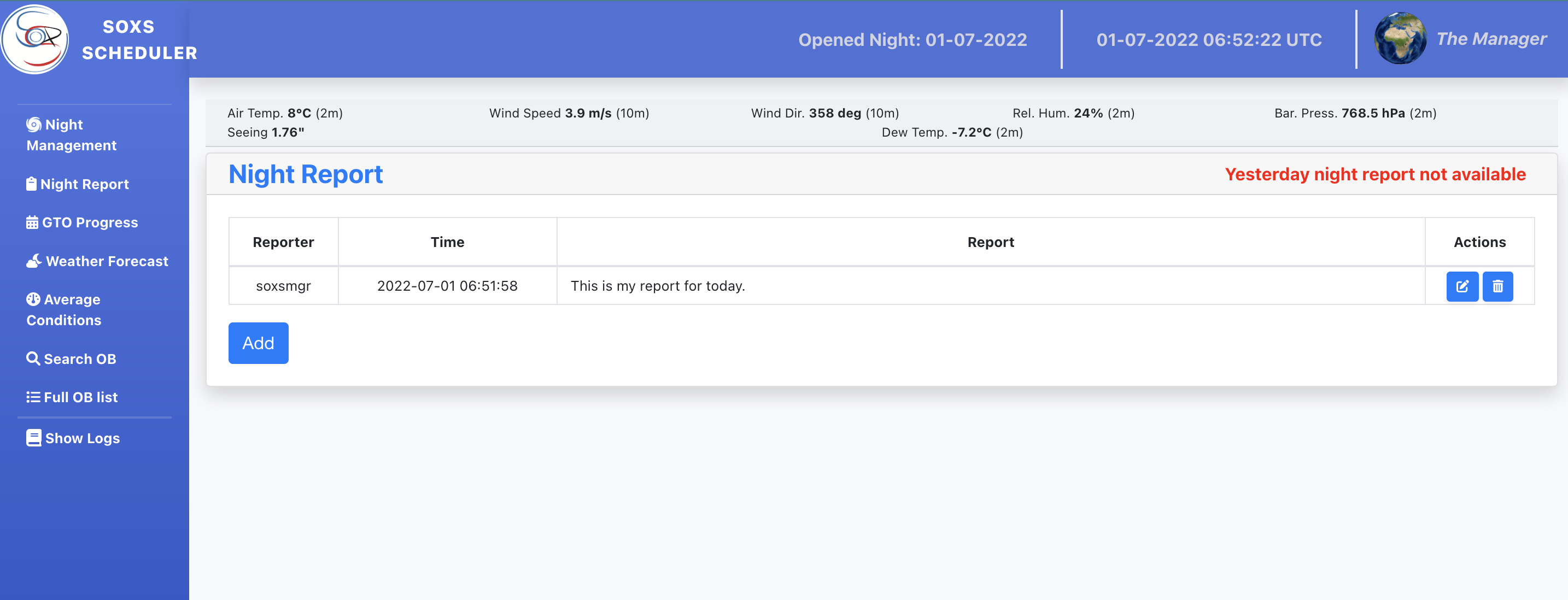}
        \includegraphics[width=0.8\textwidth]{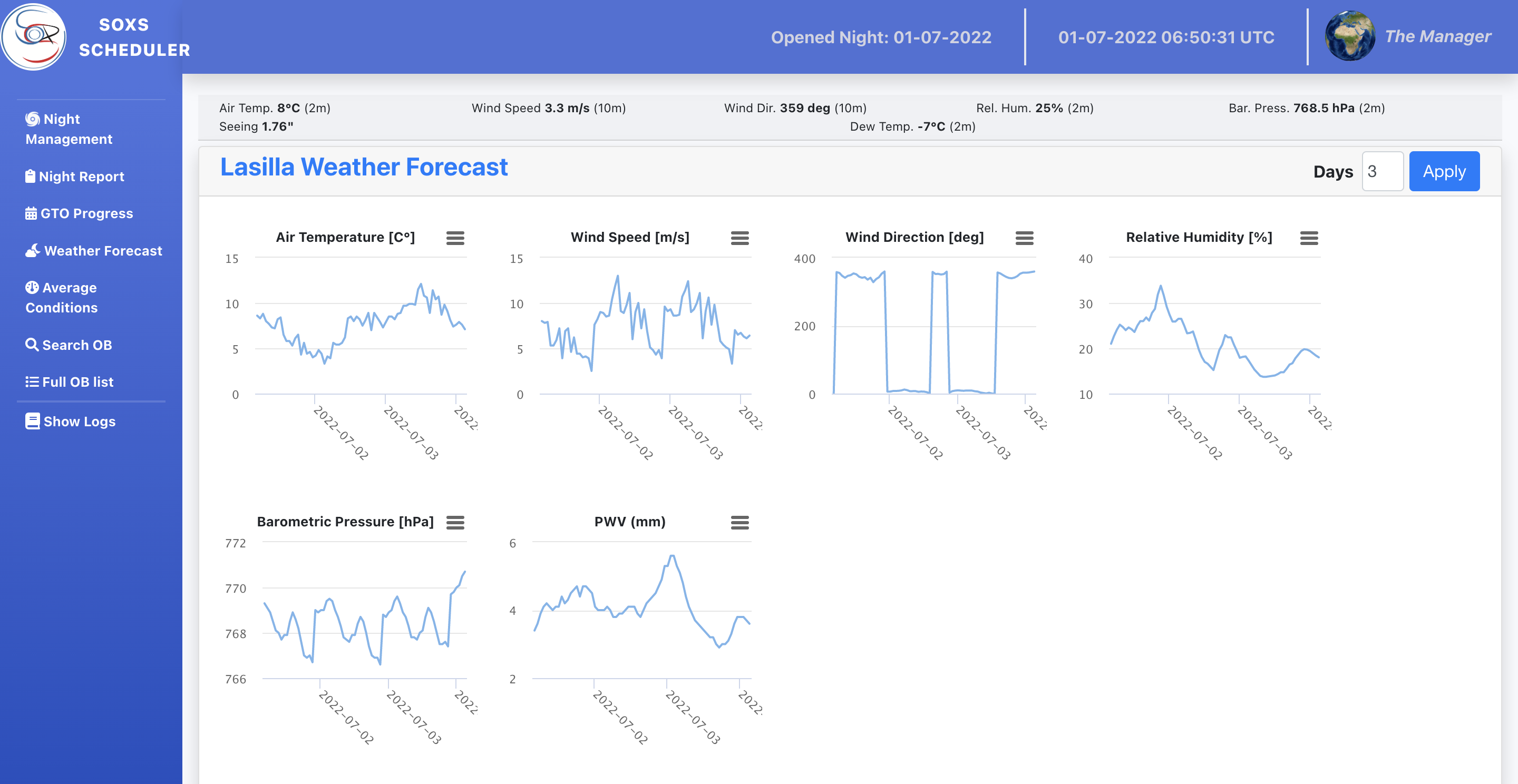}
    \caption{Top: report page. Bottom: weather forecasts in la Silla.}
    \label{fig:if02}
\end{figure}

\begin{figure} [h!]
    \centering
    \includegraphics[width=0.8\textwidth]{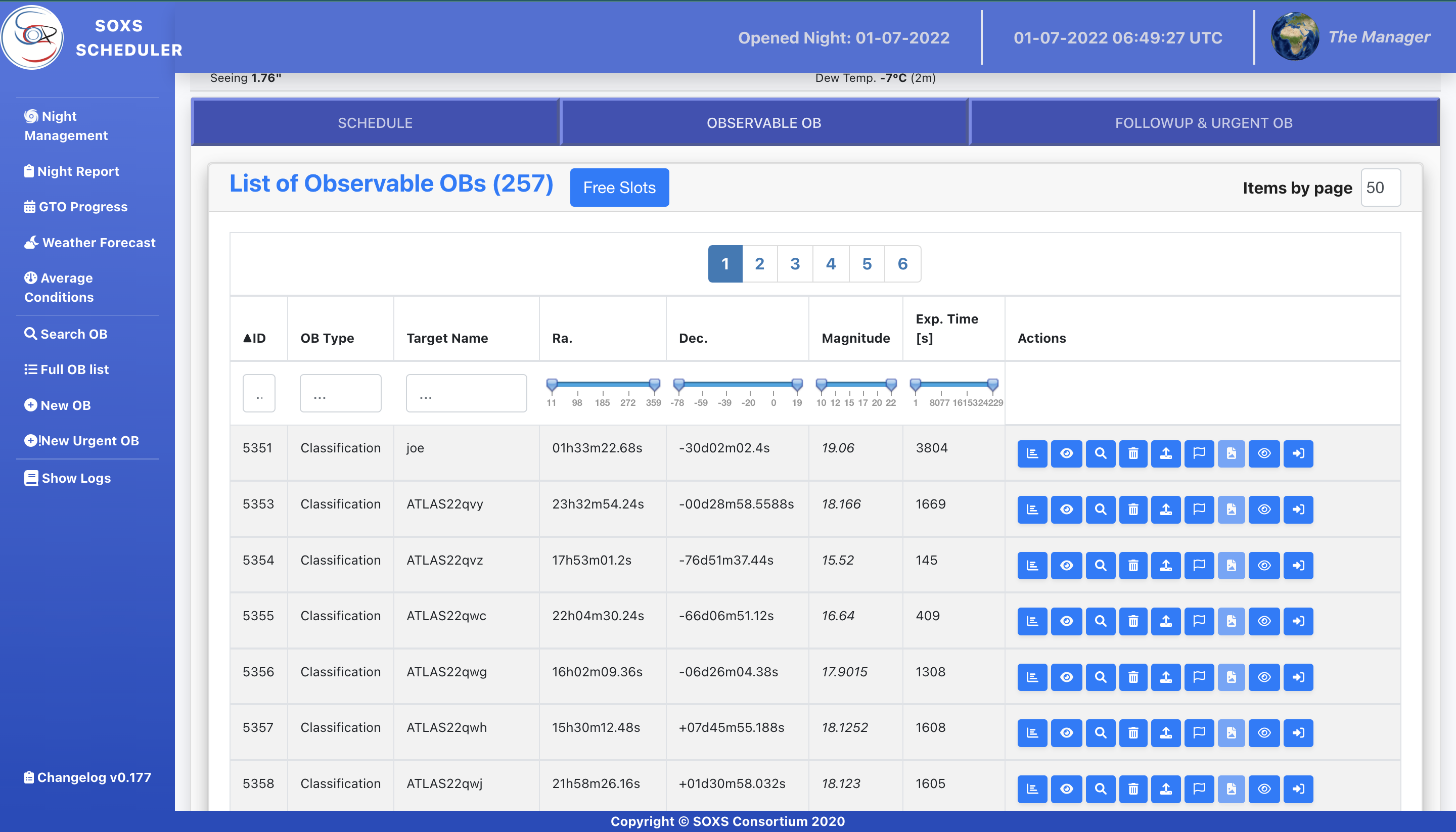}
    \caption{The page showing observable OBs.}
    \label{fig:if04}
\end{figure}

The web interface of the API has already been presented\cite{landoni2020soxs} in past papers, so we will summarize its properties and focus on its new features.
The applications is accessible to three groups of users: administrator (for internal maintenance), Science Team and Operator. Each of these user group will be able to visualize a different web page. The administrator will have access to the configuration of the application and the complete log of the application activity, while also having the ability to add or remove other users. The operator will be able to see only operator’s activity in the application logs, with the homepage showing the current observation and the completed ones. The operator will have the ability to add reports during the night, that will be visible to every user. The Science Team interface is the one that will be used during the planning of the night, and we will see it more in depth. 
The homepage (see Fig. \ref{fig:if01}) of the science team will open on the night management. The main action to perform will be to open a new night, by clicking on the relative button. A lateral menu shows: 
\begin{itemize}
\item Night management (home page Fig. \ref{fig:if01});
\item Night report, a page where the user can insert a report, read the previous ones and view or download the reports from the previous night (\ref{fig:if02}, top panel);
\item GTO progress, not implemented yet, it will show the current percentage of the GTO divided by ESO and the Consortium, possibly with its subgroups shown;
\item Weather forecasts, for the next three nights on the mountain (\ref{fig:if02}, bottom panel);
\item Average conditions, not implemented yet, will show graphs of the average observing conditions for the observations, cumulatively and separated for ESO and Consortium;
\item Search OB, a tool to search for an OB by its main identifiers.
\item Full OB list, the complete list of obs present in the database;
\item New ob, allows to create a low priority ob;
\item New urgent ob, allows to create a fully costumizable top priority target;
\item Show logs, opens the activity logs.
\end{itemize}
Further, at the very top of the page the current open night date, the current UTC time and the user name will be shown. Just below it, there will be the current weather conditions in la Silla. Below those, three tabs can be clicked, one being the working environment for the schedule, the second showing observable obs and allowing the manual scheduling (Fig. \ref{fig:if04}), the third will show only follow-up and urgent obs present in the database. 
On the ``proposed schedule" is clicked, the scheduler creates a schedule with the algorithms explained above. This is shown in chronological order. The interface shows a ``lock" button with the OB ID, which fixes the time of observation, the tool used for scheduling (Priority scheduler or gap filler algorithm), the target name, its P2 synchronization status, beginning and ending of observations. On the far right, various action buttons are available. The first one is a ``quick view", that shows basic information about the OB (the previous ones plus coordinates and slots occupied), the second one is a detailed view, that shows everything the database has on the OB. The third one shows observability information, such as the specific grades. The last action button is to substitute the ob. Upon clicking, the scheduler finds alternative OBs to fill in the gap, shows them and lets the user select one. 
\begin{figure} [h!]
    \centering
    \includegraphics[width=0.8\textwidth]{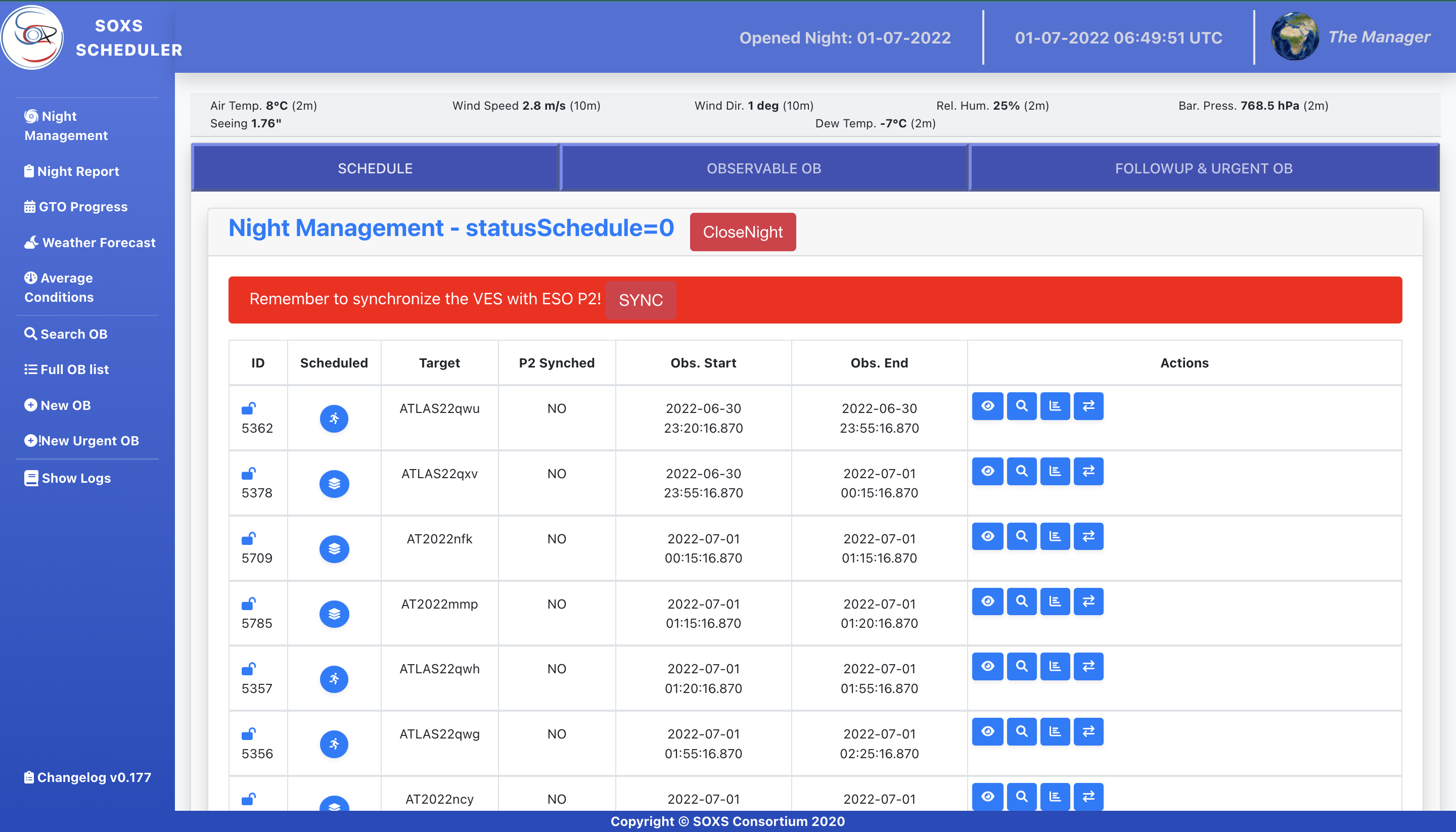}
    \caption{The proposed schedule. Each object is shown in chronological order, and from left to right it shows: a ``lock" button that fixes and unfixes OBs, with the OB ID, the target name, the status of synchronization with P2, the beginning and ending of observations. On the left, there are the action buttons that allow for review, quick and detailed of the ob, and the action of substituting it with another object.}
    \label{fig:if03}
\end{figure}

\section{Conclusions}
The scheduling system for the SOXS instrument had the necessity to bring forward many innovations in the field, both pushed by the target nature and by the new ESO policy. The development of the scheduler has achieved the required autonomy for all the main processes, such as filtering of the targets, constraint evaluation, schedule proposal. The algorithms used are flexible enough to be applied to any ground-based telescope or instrument, and their fast response time allows us to use them in many different circumstances.  Our plan for the night management calls for these already tested algorithms, and upon completion will be the first fully automated system used in ground-based observations. Future updates will implement this last part and improve the already existing ones with fine-tuning and stress-testings. Most importantly, the feedback from users in the Consortium will be considered before deployment, in order to guarantee the meeting of the needs of each community involved. 

\newpage
\bibliography{report} 
\bibliographystyle{spiebib} 
\end{document}